\documentclass[11pt]{article}
\usepackage{amssymb}
\usepackage{multicol}
\usepackage{graphicx,graphics}
\usepackage{amsfonts}
\usepackage{color}
\usepackage{subfigure}

\newcommand{\rem}[1]{}
\DeclareMathAlphabet{\mathbi}{OML}{cmm}{b}{it} 
\newtheorem{theorem}{Theorem}

\newcommand{\bx}{\mathbi{x}}

\newcommand{\bel}{\begin{equation}\label}
\newcommand{\ee}{\end{equation}}
\newcommand{\ben}{\begin{enumerate}}
\newcommand{\een}{\end{enumerate}}
\newcommand{\bit}{\begin{itemize}}
\newcommand{\eit}{\end{itemize}}
\newcommand{\I}{\int_{\mathcal{V}}}

\newcommand{\bdf}{\mathbi{f}}

\newcommand{\bu}{\mathbi{u}}

\newcommand{\bom}{\mbox{\boldmath$\omega$}}

\newcommand{\beq}{\begin{eqnarray}\label} 
\newcommand{\eeq}{\end{eqnarray}} 
\newcommand{\bc}{\begin{center}} 
\newcommand{\ec}{\end{center}} 
\newcommand\shalf{\ensuremath{{\scriptstyle\frac{1}{2}}}}

\newcommand{\Rey}{Re}

\newcommand{\Gr}{Gr}
\newcommand{\non}{\nonumber}

\makeatletter\@addtoreset{equation}{section}\makeatother

\begin{document}
\sf
\par\vspace{1mm}
\bc
\textbf{\large Estimating intermittency in three-dimensional Navier-Stokes turbulence}
\ec
\bc\textbf{\large J. D. Gibbon}\par\vspace{3mm}Department of Mathematics,
\par\vspace{1mm}Imperial College London SW7 2AZ, UK\par\vspace{1mm}
\textit{\small email: j.d.gibbon@ic.ac.uk}\\
\ec
\begin{abstract}
The issue of why computational resolution in Navier-Stokes turbulence is so hard to achieve is 
addressed. It is shown that Navier-Stokes solutions can potentially behave differently in two 
distinct regions of space-time $\mathbb{R}^{\pm}$ where $\mathbb{R}^{-}$ is comprised of a union 
of disjoint space-time `anomalies'. Large values of $|\nabla\bom|$ dominate $\mathbb{R}^{-}$, 
which is consistent with the formation of vortex sheets or tightly-coiled filaments. The local 
number of degrees of freedom $\mathcal{N}^{\pm}$ needed to resolve the regions in $\mathbb{R}^{\pm}$ 
satisfies $$\mathcal{N}^{\pm}(\bx,\,t)\lessgtr c_{\pm}\mathcal{R}_{u}^{3}$$ where $\mathcal{R}_{u} 
= uL/\nu$ is a Reynolds number dependent on the local velocity field $u(\bx,\,t)$.
\end{abstract}

\section{\sf\label{intro}\large Introduction}


The space-time distribution and morphology of the vorticity and strain fields in three-dimensional 
Navier-Stokes turbulence has remained a puzzle since Batchelor and Townsend discovered experimentally 
the phenomenon of intermittency in experimental flows. Instead of observing Gaussian behaviour in the 
flatness factor and similar quantities, they discovered the spiky spectra which are now recognized as 
typical for intermittent turbulent flows \cite{Kuo71,MS91,DCB,AT01,Lathrop,GL93,Frischbk,BMV}. 
\begin{figure}[here]
\bc\includegraphics[angle=0,scale=0.4,draft=false]{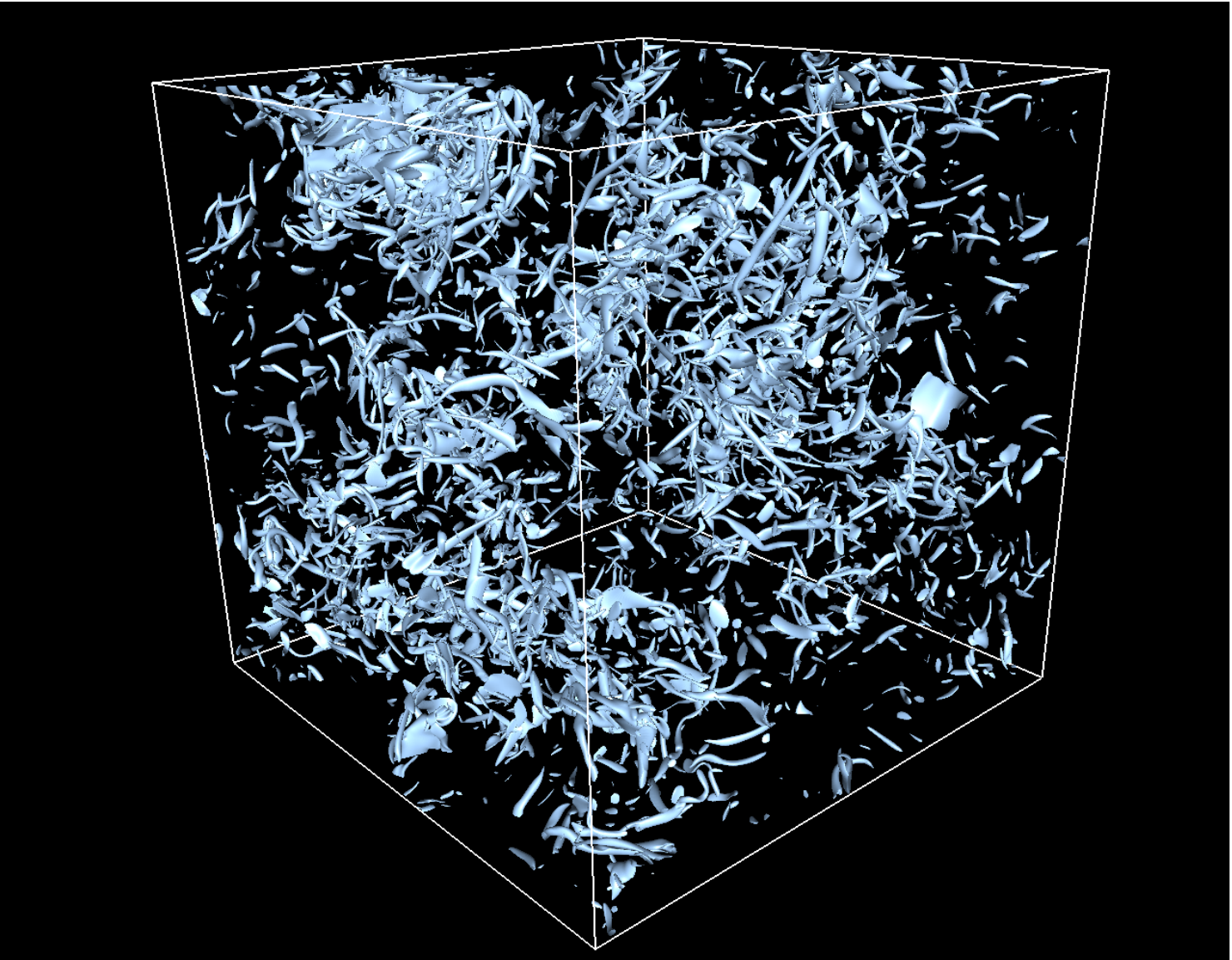}
\caption{\scriptsize\sf The picture, courtesy of J\"org Schumacher of the Technical University
Ilmenau, is a three-dimensional statistically stationary homogeneous isotropic flow at a 
Taylor microscale Reynolds number of 107 showing iso-surfaces of $\omega^2$ at a level of $10$ 
times the average of $\omega^{2}$. The cube with side-length $2\pi$ is resolved with $2048^3$ 
grid points which translates into 3 grid spacings for the Kolmogorov length $\Lambda_{k}$.}\ec
\label{fig1}
\end{figure}
\par\vspace{-3mm}\noindent
The application of colour graphics in this past generation has dramatically illustrated how the 
morphologies of the vorticity and strain fields are typically dominated by `thin sets'\,: see 
\cite{Kaneda02,LASci05}. These usually form initially as quasi-two-dimensional vortex sheets 
which, under interaction, roll up into a tangle of quasi-one-dimensional tubes \cite{VM94}. 
It is also important to note that vorticity and strain accumulate on significantly different 
sets \cite{Kerr85}: indeed, there has been some debate over their relative importance 
\cite{JWSR93,AT98,AT01}. Figure 1, courtesy of J\"org Schumacher, is a snapshot illustration 
of the vorticity (enstrophy) field . Experiments show that these structures spontaneously 
appear and disappear \cite{DCB,CDC,Emmons51}. While there exists an extensive literature on 
intermittency in the statistical physics literature concerning Kolmogorov's theory 
\cite{GL93,Frischbk,BMV}, no satisfactory theoretical explanation for the high degree of 
space-time complexity of these phenomenon has ever been given based on three-dimensional 
Navier-Stokes solutions, nor has any mathematical explanation been forthcoming why vortex 
sheets should be, at least initially, the favoured topology. The consequences of this behaviour 
is far-reaching.  The spontaneous appearance of these structures, often at very short length 
scales, creates severe resolution problems which have not been been wholly solved despite 
the increase of computing power in this past generation. To provide a partial theoretical 
explanation of this is one of the main aims of this paper. Landau's heuristic estimate for the 
number of degrees of freedom $\mathcal{N}\sim \Rey^{9/4}$ needed to resolve a turbulent flow 
is based on space-time averages and is by no means enough to resolve the thin structures 
discussed above\,: for instance see Kerr \cite{Kerr85}, Schumacher, Sreenivasan \& Yakhot 
\cite{SSY07}, Schumacher, Sreenivasan \& Yeung \cite{SSY05} and Sreenivasan \cite{Sreeni04}. 

Turning now to the status of Navier-Stokes solutions, there are generally two prevailing views. 
The first, which is generally held by the computational fluid dynamics community, is that the 
Navier-Stokes equations are regular\,: that is, it is believed that unique solutions exist that 
can ultimately be resolved provided enough computing power is made available in the future. This 
is equivalent to the assumption that the Navier-Stokes equations possess strong solutions.  The 
second view, held more by Navier-Stokes analysts, is that the unsolved regularity problem leaves 
open the possibility of 
singularity formation \cite{Leray,Lady,CF,FMRT}. Caffarelli, Kohn and Nirenberg \cite{CKN} have 
shown that the potentially singular set has zero one-dimensional Hausdorff measure, which means 
that if singularities do occur in space-time then they must be rare events. Various routes to the 
efficient construction of suitable weak solutions are in evidence \cite{Lin98,LadySeregin,ChoeLewis,He04}.  


For the purposes of providing a mathematical explanation for the resolution problem outlined 
above, the first view will be taken in this paper. In fact it cannot be emphasized enough that 
a flow may be regular but could nevertheless be highly intermittent thus rendering the singular 
set empty.  To make these arguments accessible to a wide readership the proofs, although minimal, 
are relegated to appendices which can be ignored if the reader so wishes.

\section{\sf\label{setting}\large Results based on space-time averages}

The setting is the following: we consider the incompressible ($\rm{div}\,\bu = 0$), three-dimensional 
Navier-Stokes equations for the velocity field $\bu(\bx,\,t)$ with mean-zero, divergence-free forcing,
\bel{NS1}
\bu_{t} + \bu\cdot\nabla\bu = \nu\Delta\bu - \nabla p + \bdf(\bx)\,,
\ee
on a periodic three-dimensional domain $\mathcal{V}= [0,\,L]^3$. Leray's energy inequality \cite{CF,FMRT} 
\bel{lerayenergy}
\shalf\frac{d~}{dt}\|\bu\|_{2}^{2} \leq - \nu \|\bom\|_{2}^{2} + \|\bu\|_{2}\|\bdf\|_{2}
\ee
gives some information on the energy $\|\bu\|_{2}^2$. In (\ref{lerayenergy}) $\|\cdot\|_{2}$ is the $L^2$-norm 
on the periodic volume $\mathcal{V} = [0,\,L]^3$.  The method of Doering and Foias \cite{DF02} is now
applicable in which the space-time averaged velocity $U$ and the energy dissipation rate $\varepsilon$ are 
found to be \textit{a priori} bounded quantities
\bel{int1}
U^2 = L^{-3}\left<\|\bu\|_{2}^{2}\right>\qquad\qquad 
\varepsilon = \nu L^{-3}\left<\|\bom\|_{2}^{2}\right>\,,
\ee
in which the symbol $\big<\cdot\big>$ for the long-time average is
\bel{timeavdef}
\left<g\right> = \limsup_{t\to\infty}\frac{1}{t}\int_{0}^{t}g(\tau)\,d\tau\,.
\ee
For forcing concentrated around one length scale $\ell$ with $\ell$ taken for simplicity to be 
$\ell = L/2\pi$, the Reynolds and Grashof numbers are defined by
\bel{Reydef}
\Rey = \frac{U\ell}{\nu}\,,\qquad\qquad \Gr = \frac{\ell^{3}f_{rms}}{\nu^2}\,,
\ee
in which $f_{rms} = L^{-3/2}\|\bdf\|_{2}$. (\ref{lerayenergy}) gives \cite{DF02}
\bel{leray1}
\left<\I|\bom|^{2}\,dV\right> \leq \nu^{2}L^{-1}\Gr\Rey\,,
\qquad\Rightarrow\qquad
\varepsilon \leq \nu^{3}L^{-4}\Gr\Rey\,. 
\ee
Moreover, in \cite{DF02} it has been shown that at high values of $\Gr$ Navier-Stokes solutions 
obey $\Gr \leq \Rey^2$ and so the right hand side of (\ref{leray1}) can be estimated as $\Gr\Rey 
\leq c\,\Rey^3$. Because $U$ and $\varepsilon$ are bounded quantities, the definitions of the 
respective inverse Taylor micro-scale and Kolmogorov lengths $\Lambda_{T}^{-1}$ and $\Lambda_{k}^{-1}$ 
are on a sound footing\,:
\bel{lamT}
\Lambda_{T}^{-2} = \frac{\varepsilon}{\nu\,U^{2}} = 
\frac{\left<\|\bom\|_{2}^{2}\right>}{\left<\|\bu\|_{2}^{2}\right>}\,,
\qquad\qquad 
L\Lambda_{k}^{-1} = \left(\frac{\varepsilon}{\nu^3}\right)^{1/4}\,.
\ee
Then it is easily shown that \cite{DF02}
\bel{int3}
L\Lambda_{T}^{-1} \leq c\,\Rey^{1/2}\,, 
\qquad\qquad 
L\Lambda_{k}^{-1} \leq c\,\Rey^{3/4}\,. 
\ee
This leads to an estimate for the number of degree of freedom in the system $\mathcal{N}(\Lambda_{k}) 
\leq c\,\Rey^{9/4}$ based on the number of small vortices of volume $\Lambda_{k}^3$ relative to the box 
volume $L^3$. This is consistent with Landau's heuristic and Kolmogorov's scaling arguments \cite{Frischbk,BMV}. 
These space-time averages 
are extremely useful in setting average magnitudes relative to which other quantities can be measured but they 
hide strong spiky variations in local behaviour. Two new results, which are based on an extension of (\ref{leray1}) 
to higher moments ($m\geq 1$), are proved in Appendix \ref{AppA} 
\bel{2mav} 
\left<\left(\I |\bom|^{2m}\,dV\right)^{\frac{1}{4m-3}}\right> 
\leq c_{0,m}\nu^{\frac{2m}{4m-3}}L^{-1}\Rey^3\,,
\ee
and
\bel{grad2mav} 
\left<\left(\I |\nabla\bom|^{2m}\,dV\right)^{\frac{1}{6m-3}}\right> 
\leq c_{1,m}\nu^{\frac{2m}{6m-3}}L^{-1}\Rey^3\,.
\ee
In (\ref{grad2mav}) when $m=1$ the exponent on the integral within the time average is not unity 
but 1/3.  These averages hint at some control over large fluctuations in $\bom$ but not enough 
information is available to understand the behaviour of local space-time variations. Nevertheless, 
(\ref{grad2mav}) plays an important role and will be referred to later in \S\ref{stresults}.

\section{\sf\label{stresults}\large Local space-time results}

The task is now to consider how a fluid can behave in local regions of space-time based on the 
assumption, discussed in \S\ref{setting}, that strong solutions exist. This approach requires some 
differences of definition, particularly in the Reynolds and Grashof numbers of the last section, 
$\Rey$ and $\Gr$, whose definitions were based on a space-time average in the case of $\Rey$ and 
a spatial average in the case of $\Gr$. Now we define local Reynolds and Grashof numbers as
\bel{Rdef1}
\mathcal{R}_{u}(\bx,\,t) = \frac{L|\bu(\bx,\,t)|}{\nu}\,,\quad\qquad
\mathcal{G}(\bx) = \frac{L^{3}|\bdf(\bx)|}{\nu^2}\,.
\ee
The idea is to consider the enstrophy and palenstrophy 
\bel{enst1}
H_{1}(t) = \I|\bom|^{2}\,dV\qquad\quad
H_{2}(t) = \I|\nabla\bom|^{2}\,dV\,.
\ee
and derive, as in Appendix \ref{AppB}, a differential inequality involving these 
(valid for $6 \geq q > 3$)
\bel{H1}
\shalf\dot{H}_{1} \leq -\frac{(q-3)}{2(q+6)}\,\nu H_{2} + 
c_{q}\,\nu^{-\frac{9}{q-3}}\|\bu\|_{q}^{\frac{3q}{q-3}} + 
\frac{q+6}{2(q-3)}\nu^{-1}\|\bdf\|_{2}^{2}\,.
\ee
This is a modification of a version first proved by Ladyzhenskaya \cite{Lady} in a regularity 
proof conditional upon the velocity $\bu$ being bounded in $\|\bu\|_{q}$ for $q>3$.  Instead, 
when $q=6$, the nonlinear term is proportional to the perfect integral $\|\bu\|_{6}^{6}$. 
Time averaging as in (\ref{timeavdef}) and re-scaling so that each term is dimensionless, 
(\ref{H1}) converts to a 4-integral 
\bel{np6}
\left<\I\left\{c_{6}\mathcal{R}_{u}^{6} - 
\frac{L^{2}}{8\omega_{0}^{2}}|\nabla\bom|^{2} + 2\mathcal{G}^{2}\right\}\,dV\right> 
\geq 0\,.
\ee
A trivial but important observation is that a positive integral need not necessarily have an 
integrand that is positive everywhere in space-time. Some consequences of this are\,:
\ben
\item There are regions of space-time $\mathbb{R}^{+} \subset \mathbb{R}^{4}$ on which 
\bel{con1}
\frac{L^{2}}{8\omega_{0}^{2}}|\nabla\bom|^{2}  \leq 
c_{6}\mathcal{R}_{u}^{6} + 2\mathcal{G}^{2}\,.
\ee

\item In contrast, there are (potential) disjoint regions of, or `anomalies', in space-time on which 
\bel{con2}
\frac{L^{2}}{8\omega_{0}^{2}}|\nabla\bom|^{2}  > 
c_{6}\mathcal{R}_{u}^{6} + 2\mathcal{G}^{2}\,.
\ee
The union of these anomalies we call $\mathbb{R}^{-} \subset \mathbb{R}^{4}$. The contribution from 
$\mathbb{R}^{-}$ cannot be too large given the positivity of the 4-integral in (\ref{np6}).

\item Given that the set $\mathbb{R}^{-}$ is non-empty, \textit{the first observation is that the 
very large nonlinearity $\mathcal{R}_{u}^{6}$ amplifies the response in the magnitude of 
$|\nabla\bom|^2$ to relatively small fluctuations in the local velocity field $\bu$}. Using the 
fact that the higher moments of $|\nabla\bom|$ are controlled as in (\ref{grad2mav}), it is clear 
that $|\nabla\bom|$ cannot become too large everywhere in $\mathbb{R}^4$ or this average will be 
violated. Very large behaviour in $\mathbb{R}^{-}$ must therefore be balanced by smaller behaviour 
in $\mathbb{R}^+$.  

The product of this amplified and highly uneven response is an intermittent spectrum.
This is consistent with the remark of Batchelor and Townsend \cite{BT49} where they suggested that 
large wave-number components are concentrated in isolated flow regions with an uneven energy 
distribution associated with the small scale components. 

\item \textit{The second observation is that the sudden rapid increase of the gradient $|\nabla\bom|$ 
as one moves from a region in $\mathbb{R}^{+}$ across into an anomaly in $\mathbb{R}^{-}$ is consistent 
with the formation of vortex sheet-like structures\footnote{\sf Formally, a vortex sheet is considered 
to have formed when a jump occurs in tangential vorticity as one moves in a normal direction\,; in 
our case we are unable to make any distinction between the normal and tangential directions. See 
Majda \& Bertozzi \cite{MB01} for a discussion and references on the formation of vortex sheets in 
the Euler equations.} or perhaps tightly-coiled filaments}. The subsequent roll-up of these sheets 
when they interact, as observed in numerical experiments, is not explained but since both topologies 
have a small packing fraction the roll-up of one into the other would not change this.

\item Defining a local Kraichnan length as 
\bel{kr1}
\big(L\lambda_{kr}^{-1}\big)^{6} = \frac{|\nabla\bom|^{2}}{\nu^2} =
\frac{L^{2}|\nabla\bom|^{2}}{\omega_{0}^2}\,,
\ee
then (\ref{con1}) and (\ref{con2}) can be re-written as 
\bel{kr2}
\big(L\lambda_{kr}^{-1}\big)^{6} \lessgtr
8c_{6}\mathcal{R}_{u}^{6} + 16\mathcal{G}^{2}\qquad\mbox{on}\qquad \mathbb{R}^{\pm}\,.
\ee
Thus, to resolve the structures in $\mathbb{R}^{\pm}$, the number of degrees of freedom$\mathcal{N}^{\pm}$ 
is estimated as  
\bel{kr3}
\mathcal{N}^{\pm} \lessgtr c_{\pm}\big(\mathcal{R}_{u}^{3} + \mathcal{G}\big)
\qquad\mbox{on}\qquad \mathbb{R}^{\pm}
\ee
compared with the standard $\Rey^{9/4}$ needed on the space-time average. Moreover, $\mathcal{R}_{u}$ 
in particular, might be considerably larger than $\Rey$ in or near some anomalies.  Given that (\ref{kr3}) 
is a lower bound in $\mathbb{R}^{-}$, $\mathcal{N}^{-}$ might be considerably larger than $\mathcal{R}_{u}^{3}$, 
showing why resolution might be lost locally, if $u$ becomes too large. 
\een

\section{\sf\label{discuss}\large Summary and discussion\,: vorticity versus strain}

The arguments of the previous section show clearly that for strong solutions of the Navier-Stokes 
equations, space-time can potentially be split into two parts $\mathbb{R}^{\pm}$. While the closed 
set $\mathbb{R}^{+}$ is dominant because of the positivity of the space-time integral in (\ref{np6}), 
the vary lare lower bound $\mathcal{R}_{u}^{6}$ in (\ref{kr2}) provokes high values of $|\nabla\bom|$ 
in the space-time anomalies, whose union is the open set $\mathbb{R}^{-}$. The control of space-time 
averages of higher moments in (\ref{grad2mav}) insists that large values of $|\nabla\bom|$ in 
$\mathbb{R}^{-}$ must be balanced by smaller values in $\mathbb{R}^{+}$. The violent increase of 
$\bom$ in the relatively small space-time region occupied by these anomalies suggests the formation 
of vortex sheet-like structures or tightly-coiled filaments. Numerical evidence suggests that sheets 
roll up into tubes when they interact, as in Figure 1 \cite{VM94}. Indeed, both topologies have an 
appropriately small packing fraction which is no doubt why they are commonly observed. The bounds 
on $\mathcal{N}^{\pm}$ in (\ref{kr3}) are dependent on the velocity field at local space-time points 
of the flow. Despite the assumption of regularity of solutions, $\mathcal{R}_{u}$ may potentially 
reach large values, making the $\mathcal{R}_{u}^{3}$ lower bound enormous.  This would account for 
local difficulties in resolution and illustrates the need to carefully monitor values of 
$\mathcal{R}_{u}$ in a numerical calculation\,: see, for instance, \cite{SSY07}. 

These ideas can also be used in an alternative manner to see the effect of strain\,: see 
\cite{JWSR93,AT98,AT00,AT01} and \cite{Kerr01}. In this case it is more appropriate to define different 
local Reynolds numbers as
\bel{Redef}
\mathcal{R}_{\omega} = \frac{L^{2}|\bom|}{\nu}\,,\qquad\qquad
\mathcal{R}_{\rho} = \frac{L^{2}\rho_{s}}{\nu}\,.
\ee
In (\ref{Redef}) $\rho_{s}(\bx,\,t)$ is the spectral radius of the strain rate matrix $S$ which appears
because $\bom\cdot\bom\cdot\nabla\bu = \bom\cdot S\bom$. The equivalent of the 4-integral in (\ref{np6}) 
is
\bel{Scalcn}
\left<\I\left\{2\mathcal{R}_{\rho}\mathcal{R}_{\omega}^{2} - 
\frac{L^{2}}{\omega_{0}^{2}}|\nabla\bom|^{2} + \mathcal{G}^{2}
\right\}\,dV\right> \geq 0\,.
\ee
Similar conclusions can be reached to those of \S\ref{stresults} regarding the effect the strain field 
and vorticity fields have on $|\nabla\bom|$ in regions $\mathbb{R}^{\pm}_{s}$ of space-time. These 
will be different from those regions contained in $\mathbb{R}^{\pm}$ of (\ref{np6}). An estimate for 
the number of degrees of freedom needed to resolve $\mathbb{R}^{\pm}_{s}$ -- the equivalent of 
(\ref{kr3}) -- is 
\bel{krs1}
\mathcal{N}^{\pm}_{s} \lessgtr c^{\pm}_{s}\,\left\{\mathcal{R}_{\rho}^{1/2}\mathcal{R}_{\omega} + 
\mathcal{G}\right\}\,.
\ee

\par\vspace{2mm}\noindent
\textbf{Acknowledgements\,:} My thanks are due to Panagiota Daskalopoulos, Charles Doering, Raymond
Hide, Darryl Holm, Bob Kerr, Gerald Moore, Trevor Stuart, Edriss Titi, J\"org Schumacher and 
Arkady Tsinober for discussions.

\appendix
\section{\sf\label{AppA}\large Proof of (\ref{2mav}) and (\ref{grad2mav}) in \S\ref{setting}}

The estimates in this appendix are based on the assumption that strong solutions of the Navier-Stokes
equations exist. Consider the result of Foias, Guillop\'e and Temam \cite{FGT} for time averages of 
higher semi-norms which is deliberately written for our present purposes in terms of $\Gr$ and $\Rey$ 
\bel{Hntimav1}
\left<H_{n}^{\frac{1}{2n-1}}\right> \leq 
c_{n}\nu^{\frac{2}{2n-1}}L^{-1}a_{\ell}^{4}\Gr\Rey\,.
\ee
A Sobolev inequality for $m > 1$ gives 
\bel{Omsob1}
\|\bom\|_{2m} \leq c_{m}\|\nabla^{2}\bom\|_{2}^{a}\|\bom\|_{2}^{1-a} 
\ee
where $a = 3(m-1)/4m$.
Thus, taking $n = 3$ in (\ref{Hntimav1}), we have 
\beq{Omsob2}
\left<\|\bom\|_{2m}^{\frac{2m}{4m-3}}\right> &\leq& 
c_{m,1}\left<\left(H_{3}^{1/5}\right)^{\frac{15(m-1)}{4(4m-3)}}H_{1}^{\frac{m+3}{4(4m-3)}}
\right>\nonumber\\
&\leq& c_{m,1}\left<\left(H_{3}^{1/5}\right)\right>^{\frac{15(m-1)}{4(4m-3)}}
\left<H_{1}\right>^{\frac{m+3}{4(4m-3)}}\nonumber\\
&\leq& c_{m,1}\nu^{\frac{2m}{4m-3}}L^{-1}a_{\ell}^{4}\Gr\Rey\,.
\eeq
The averages of $|\nabla\bom|^{2m}$ in (\ref{grad2mav}) can be found in the same 
way using $n=4$ in (\ref{Hntimav1}). 

\section{\sf\label{AppB}\large Proof of Ladyzhenskaya's inequality (\ref{H1})}

Consider $H_{1}$ defined in (\ref{enst1}). For strong solutions
\bel{H1in}
\dot{H}_{1} = \I \bom\cdot\left\{\nu\Delta\bom + \bom\cdot\nabla\bu + \mbox{curl}\,\bdf\right\}\,dV\,.
\ee
Using the divergence theorem to estimate the vortex stretching term we find
\bel{np1}
\left|\I \bom\cdot\bom\cdot\nabla\bu\,dV\right| \leq \|\nabla\omega\|_{2}\|\omega\|_{s}\|\bu\|_{q}
\ee
where $q^{-1} + s^{-1} = \shalf$. A Sobolev inequality on $\|\omega\|_{s}$ gives
\bel{np2a}
\|\omega\|_{s} \leq \|\nabla\bu\|_{s} \leq c\,\|\nabla^{2}\bu\|_{2}^{a}\|\bu\|_{q}^{1-a}
\ee
where $a$ is given by $a = \frac{12-q}{6+q}$.  There are restrictions on $a$ in the form 
$\frac{1}{2} \leq a < 1$, which means that $3 < q \leq 6$. Thus we have 
\beq{np3}
\left|\I \bom\cdot\bom\cdot\nabla\bu\,dV\right| &\leq &
c\,\|\nabla\omega\|_{2}^{1+a}\|\bu\|_{q}^{2-a}\non\\
&= & \left(\nu H_{2}\right)^{\frac{1+a}{2}}
\left(c\,\nu^{-\frac{1+a}{1-a}}\|\bu\|_{q}^{\frac{2(2-a)}{1-a}}\right)^{\frac{1-a}{2}}\non\\
&\leq& \frac{1}{2}(1+a)\nu H_{2} + c_{q}\,\nu^{-\frac{1+a}{1-a}}\|\bu\|_{q}^{\frac{2(2-a)}{1-a}}\,.
\eeq
(\ref{H1in}) becomes
\bel{np4}
\shalf \dot{H}_{1} \leq - \nu\left(\frac{q-3}{q+6}\right)H_{2} + 
c_{q}\,\nu^{-\frac{9}{q-3}}\|\bu\|_{q}^{\frac{3q}{q-3}} + 
\|\bdf\|_{2}H_{2}^{1/2}\,.
\ee
We also know that
\bel{Yinequa2}
\left[\left(\frac{q-3}{q+6}\right)\nu H_{2}\right]^{1/2}
\left[\left(\frac{q+6}{q-3}\right)\nu^{-1}\|\bdf\|_{2}^{2}\right]^{1/2} \leq
\left(\frac{q-3}{2(q+6)}\right)\nu H_{2} + 
\left(\frac{q+6}{2(q-3)}\right)\nu^{-1}\|\bdf\|_{2}^{2}
\ee
Together (\ref{np4}) gives
\bel{np5}
\shalf \dot{H}_{1} \leq -\frac{(q-3)}{2(q+6)}\,\nu H_{2} + 
c_{q}\,\nu^{-\frac{9}{q-3}}\|\bu\|_{q}^{\frac{3q}{q-3}} + 
\frac{q+6}{2(q-3)}\nu^{-1}\|\bdf\|_{2}^{2}\,,
\ee
where one recalls that $6 \geq q > 3$. This is (\ref{H1}).


\par\vspace{-3mm}

\end{document}